\documentstyle[12pt,epsfig]{article}

\begin{document}
\begin{center}
{\small \it ECOSS-18, Vienna, September 21-24, 1999, submitted to Surface 
Science\\
The abstract ID: A9XSGW}\\
\vspace*{10mm}

{\large{\bf
Submonolayer growth with decorated island edges}}
\vspace*{2mm}

Miroslav Kotrla
\footnote{Corresponding author: M. Kotrla,
Institute of Physics ASCR,
Na Slovance 2,
182~21 Praha 8, Czech Republic, tel.: +420 2 6605 2904; 
FAX +420 2 8588605, E-mail: kotrla@fzu.cz.
}

{\em Institute of Physics 
Academy of Sciences of the Czech Republic,\\
Na Slovance 2,
182~21 Praha 8, Czech Republic}
\vspace*{5mm}

Joachim Krug
\vspace*{2mm}

{ \em
Fachbereich Physik, Universit\"{a}t GH Essen, 45117 Essen,
Germany}
\vspace*{5mm}

Pavel \v{S}milauer
\vspace*{2mm}

{\em
Institute of Physics 
Academy of Sciences of the Czech Republic,\\
Cukrovarnick\'a 10, 
162~53 Praha 6, Czech Republic}

\date{\today}
\end{center}

\begin{abstract}
We study the dynamics of island nucleation in the presence of adsorbates
using kinetic Monte Carlo simulations of a two-species
growth model. Adatoms (A-atoms) and impurities (B-atoms)
are codeposited, diffuse and aggregate subject to attractive
AA- and AB-interactions. Activated exchange of adatoms with impurities
is identified as the key process to maintain decoration of island edges
by impurities during growth. While the presence of impurities 
strongly increases
the island density, a change in the scaling of island density with flux,
predicted by a rate equation theory for attachment-limited
growth [D. Kandel, Phys. Rev. Lett. {\bf 78},
499 (1997)], is not observed. We argue that, within the present model,
even completely covered island edges do not provide efficient barriers
to attachment.
\end{abstract}
{\normalsize Keywords: 
Monte Carlo simulations, Clusters, Epitaxy, Growth, Nucleation,
Surface diffusion

\newpage

The effect of impurities on crystal growth has been a long-standing
concern in surface science \cite{cabrera58}. Recent interest in this
subject has been fueled by the prospect of using
adsorbates as ``surfactants'' to improve the quality
of epitaxially grown films \cite{kandel99}.
A key mechanism which dramatically increases the (detrimental or
beneficial) effects of adsorbates is their tendency to decorate
the step edges by preferentially attaching there
\cite{vanderVegt92,kalff98}.
If this induces an additional energy barrier which
an adatom has to
overcome to be incorporated at the step edge,  
the scaling
exponent $\chi$ in the relation   
\begin{equation}
\label{N}
N \sim (F/D)^\chi
\end{equation}
between the island density $N$, the deposition flux $F$ and 
the adatom diffusion
coefficient $D$ may be affected. 
Assuming a critical (= largest unstable) island
size $i^\ast$, standard rate equation theory yields 
in two dimensions 
the expression \cite{stoyanov81,venables84}
\begin{equation}
\label{chi}
\chi = \frac{i^\ast}{i^\ast+2}
\end{equation} 
while strong incorporation barriers imply \cite{kandel97}
\begin{equation}
\label{chi2}
\chi = \frac{2 i^\ast}{i^\ast + 3}.
\end{equation}
Thus 
inspection of the flux dependence of the island density
could reveal whether or not the island edges are efficiently 
{\em passivated\/} by the adsorbates 
provided the critical island size is only slightly changed by
the presence of adsorbates. 
Markov \cite{markov97} has
pointed out that the change from (\ref{chi}) to 
(\ref{chi2}) can be understood as a transition from a
{\em diffusion limited} to a {\em kinetically limited} growth
regime.   

In this paper we report on an extensive numerical study of a two-species
growth model aimed at answering the following two questions:
First, what kinetic processes are necessary to generate an
island morphology in which the impurities decorate
the island edges? Second, how does the presence
of impurities affect the island density, and what role does
the passivation of island edges play in this context? 

We employ a 
full diffusion  
solid-on-solid model of epitaxial growth
with two surface species, which we denote $A$ and $B$.
We suppose that particles of type $A$ correspond to the growing material 
and particles of type $B$
represent the impurities.
Simulations start on a flat substrate composed only of $A$ atoms. 
Both species are randomly deposited with generally different fluxes
$F_A$ and $F_B$.
Here, we restrict ourselves to codeposition with $F_A=F_B$.
Similar results for predeposition of adsorbates (B atoms) as well as 
a study of the effect of the concentration of adsorbates will be presented
elsewhere.
The migration of each surface atom is modeled as a nearest--neighbor hopping
process with rate $R_D=k_0 \exp (-E_D /k_B T)$,
where $k_0= 10^{13}$ Hz is an adatom vibration frequency,
$E_D$ is the hopping barrier, $T$ is the substrate temperature
and $k_B$ is Boltzmann's constant.
The hopping barrier is the sum of a term from the substrate
$E_{\rm sub}$ and a contribution from each lateral nearest neighbor 
$E_{\rm n}$. 
Contributions depend on local composition: For each term we have the
four possibilities $AA$, $AB$, $BA$ and $BB$.
Thus, the hopping barrier of adatom $X$ ($A$ or $B$) is
\begin{equation}
\label{E_D^X}
E_D^X= \sum_{Y=A,B} \left( n_0^Y E_{\rm sub}^{XY}
+ n_1^{XY} E_{\rm n} ^{XY} \right) ,
\end{equation}
where $E_{\rm sub}^{XY}$ is the hopping barrier for a free $X$ adatom 
on a substrate atom $Y$, $n_0^Y$ is equal to one if the substrate
atom is of type $Y$ and zero otherwise,
$n_1^{XY}$ is the number of nearest-neighbor
$X$-$Y$ pairs, and $E_{\rm n}^{XY}$
is the corresponding contribution to the barrier (symmetric in $X$ and $Y$).
Lateral interactions between impurity atoms are neglected
($E_{\rm n}^{BB}=0$).

In the simulations reported here we used 
$E_{\rm sub}^{AA}=0.8$ eV,
$E_{\rm sub}^{AB}=0.1$ eV,
$E_{\rm sub}^{BA}=1.0$ eV,
$E_{\rm sub}^{BB}=0.1$ eV, 
and the substrate temperature $T=500$~K.
Other parameters for lateral interactions were varied.
The system sizes ranged from 300$\times$300 to 500$\times$500.
The low values of $E_{\rm sub}^{AB}$ and $E_{\rm sub}^{BB}$ ensure that
atoms deposited on top of an impurity immediately descend to the 
substrate.
We use fluxes 
in the interval from 0.00025 ML/s to 0.25 ML/s.
In the case of homoepitaxy,
different values of $E_{\rm n}^{AA}$  correspond 
to  different sizes of critical nucleus $i^\ast$.
There are two different situations
$E_{\rm n}^{AA}  > E_{\rm n}^{AB}$ and
$E_{\rm n}^{AA}  < E_{\rm n}^{AB}$.
In equilibrium at a low temperature,  
 the former case leads to the
formation of islands composed inside mainly of $A$ atoms
with $B$ atoms bounded near the edges,
whereas in the latter case it is energetically more 
favorable when $B$ atoms are inside the island.
We observed, however, 
that for our parameters, growth always leads to intermixing
of $A$ and $B$ atoms,
in both cases $E_{\rm n}^{AA}  > E_{\rm n}^{AB}$ and
$E_{\rm n}^{AA}  < E_{\rm n}^{AB}$.

Thus the energetic bias 
favoring segregation is not sufficient to obtain
configurations with impurities mostly
at island edges. To achieve this, we have to 
introduce an additional thermally activated process
of exchange of an $A$ atom approaching an island 
edge covered with an impurity. 
This is a similar process to that introduced
in Ref. 
\cite{smilauer98}
for simulation of homoepitaxy on Si(001) with predeposited
hydrogen. 
Here, we allow the exchange of an $A$ atom with an impurity 
when it has 
before the exchange process
at most one bond to another $A$ atom
in a nearest-neighbor position. 
The rate of the exchange process is set to 
$k_{\rm ex}=k_0 \exp (- E_{\rm ex}/k_B T)$, 
where the activation barrier $E_{\rm ex}$ is taken
to be independent of whether or not the A-atom has 
a nearest neighbor bond. The exchange barrier  $E_{\rm ex}$ was varied
from 0.8 eV (the diffusion barrier of free adatoms)
to 2 eV. 
For low $E_{\rm ex}$ the impurities float on the island edges, 
whereas for large $E_{\rm ex}$ they  
tend to get trapped inside an island. 
In order to obtain decorated island edges, $E_{\rm ex}$ has to be lower
than a certain value which in the present case is about 1.2 eV;
in the following we set $E_{\rm ex} = 1$ eV.

In Fig. 1a and 1b we 
show examples of typical configurations with the same partial coverage 
of both species $\theta_A = \theta_B = 0.1$ ML (i.e. total coverage
$\theta = \theta_A + \theta_B  = 0.2$ ML)
obtained by 
codeposition
with fluxes $F_A=F_B = 0.004$ ML/s.
They illustrate the effect of varying the relation between
$E_{\rm n}^{AA}$ and $E_{\rm n}^{AB}$ for $E_{\rm n}^{AA} = 0.3$ \,eV
(a typical configuration for homoepitaxy in shown in Fig. 1c).
For $E_{\rm n}^{AB} = 0.2$ \,eV some gaps can be seen in the impurity
layer surrounding the islands, while for $E_{\rm n}^{AB} = 0.4$ \,eV
the decoration is complete.
If we assume that the decorated island edge is in equilibrium with an
ideal impurity gas of coverage $\theta_B$, then 
a simple detailed balance argument shows that the concentration of
uncovered edge sites is
$(1 + \theta_B e^{E_{\rm n}^{AB}/k_B T})^{-1}$. 
Thus for $T = 500$ K, $\theta_B = 0.1$ 
and $E_{\rm n}^{AB} = 0.2$ \,eV about one tenth of the
edge sites are uncovered, in accord with the visual inspection
of Figure 1.
Correspondingly, simulations
with $E_{\rm n}^{AB} = 0.1$ \,eV show that large portions of the step edges
remain uncovered.

It can be seen from 
Figure 1
that the
island density $N$ increases with increasing $E_{\rm n}^{AB}$.
A quantitative evaluation of the dependence of island density
on flux, coverage and $E_{\rm n}^{AB}$ is given in Figure 2.
For comparison we show also results for homoepitaxial growth without
impurities. We would like to point out three noteworthy features
of these data. First, increasing
the interaction energy between adatoms and impurities dramatically
increases the island density. Second, the increase is essentially
independent of flux, i.e. the scaling exponent $\chi$
 in Eq.(\ref{N})
{\em is hardly affected by the impurities}; we find
$\chi=0.54$ for $E_{\rm n}^{AB} = 0.2$,
$\chi=0.45$ for $E_{\rm n}^{AB} = 0.4$, and
$\chi=0.54$ for homoepitaxial growth. 
Third, for large fluxes
and large values of $E_{\rm n}^{AB}$, the island density continues to
increase with coverage at least up to $\theta = 0.2$ ML. 

While a detailed discussion of these effects will be left to an
extended publication, here we provide a simple argument to 
explain why the scaling exponent (\ref{chi2}) predicted for
passivated islands is not observed in our simulations. 
The mechanism described by Kandel \cite{kandel97} relies on the
rate of attachment of an adatom to an island, $S$, to be much smaller
than the diffusion rate $D$ on the terrace. It is useful to introduce
the probability $p = S/(D + S) \approx S/D$ for an adatom to attach
during an encounter with an island edge. In the kinetically
limited growth regime, characterized by (\ref{chi2}), an adatom typically
has to visit many islands before being captured. To see when
this is the case, consider an adatom diffusing through an array of islands
of density $N$ and linear size $R \sim \sqrt{\theta_A/N}$.
During $n$ diffusion steps the adatom explores a region of area
$\sim n$, which contains $\sim N n$ islands and $\sim N n R$ edge sites.
The adatom is captured when $N n R p \approx 1$. It follows that the 
condition for encountering many islands prior to capture reads
\begin{equation}
\label{capture}
p \ll R^{-1} \sim \sqrt{N/\theta_A}.
\end{equation} 
For a given value of $p$ this places a lower bound on the island density
(and, according to (\ref{N}), a lower bound on the flux) beyond which
a crossover to conventional, diffusion limited scaling will occur.

Let us estimate $p$ for a completely decorated
island. An adatom approaching the island has two options: It can return
to the terrace with rate $k_0 e^{-(E_{\rm sub}^{AA} + E_{\rm n}^{AB})/k_B 
T}$,
or it can exchange with the impurity (and thus join the island) with rate
$k_0 e^{-E_{\rm ex}/k_B T}$. The capture probability $p$ is the probability
that the second process occurs before the first, and is given by
\begin{equation}
\label{p}
p = [1 + e^{(E_{\rm ex} - E_{\rm sub}^{AA} - E_{\rm n}^{AB})/k_B T}]^{-1}.
\end{equation}
With the standard values $E_{\rm ex} = 1$ eV, $E_{\rm sub}^{AA} = $ 0.8 
eV 
and $E_{\rm n}^{AB} \geq 0.2$ eV it can be seen that the condition
(\ref{capture}) is never satisfied.
To significantly decrease $p$, one would have to either increase
the exchange barrier $E_{\rm ex}$ relative to the diffusion barrier
$E_{\rm sub}^{AA}$, which leads to intermixing of impurities and
adatoms within the island, or to decrease the AB-interaction strength
$E_{\rm n}^{AB}$, which would increase the number of gaps in the impurity
layer and invalidate our assumption of complete decoration.
In fact the appearance of $E_{\rm n}^{AB}$ in the exponent of (\ref{p})
implies that even completely decorated island edges are
not efficiently passivated in our model, 
because the B-atoms covering the edge still
have ``bonds'' available with which to attract A-atoms and keep them
near the edge long enough for an exchange to occur. 

In conclusion, we have shown that the attachment-limited growth regime
characterized by the scaling exponent (\ref{chi2}) is not easily
realizable in a two-species growth system with isotropic
nearest neighbor interactions. It remains to be understood what 
microscopic mechanism, other than passivation of island edges, is
responsible for the strong increase in island density shown in 
Figure 2.
The most obvious possibility is that the impurities
reduce the mobility of adatoms by temporarily trapping them, thus effectively
reducing the diffusion coefficient $D$ in (\ref{N}). A detailed analysis
of this mechanism will be presented elsewhere. 

\vspace{0.5cm}

{\bf Acknowledgements.} This work was supported by Volkswagen-Stiftung
and by the COST project P3.130.

\newpage
\begin{center}
{\large
{\bf 
Figure Captions
}}
\end{center}

Fig. 1:
Examples of  configurations for 
flux $F_A = 0.004$ ML/s, adatom interaction energy
$E_{\rm n}^{AA} = 0.3$ \,eV, and
coverage $\theta_A = 0.1$ ML: 
(a) codeposition with $F_B = F_A$ and 
$E_{\rm n}^{AB} = 0.2$ \,eV,
(b) codeposition with $F_B = F_A$ and
$E_{\rm n}^{AB} = 0.4$ \,eV,
(c) homoepitaxy ($F_B = 0$).
We show only $50 \times 50$ sections of a larger simulation
box.
\vspace*{5mm}

Fig. 2:
Averaged island density as function of flux
$F_A$ for several values of the total coverage 
$\theta = \theta_A + \theta_B$ and different  energy barriers:
$E_{\rm n}^{AB} = 0.2 $ \,eV - open symbols,   
$E_{\rm n}^{AB} = 0.4 $ \,eV - filled symbols.
The adatom interaction energy
$E_{\rm n}^{AA} = 0.3$ \,eV and  
the exchange barrier $E_{\rm ex}= 1$ \,eV
are fixed, and the impurity flux $F_B = F_A$. 
The behavior in the absence of impurities
(homoepitaxy, $F_B = 0$) is shown for comparison.

\newpage
\begin{figure}[hb]
\centering
\vspace*{70mm}
\includegraphics{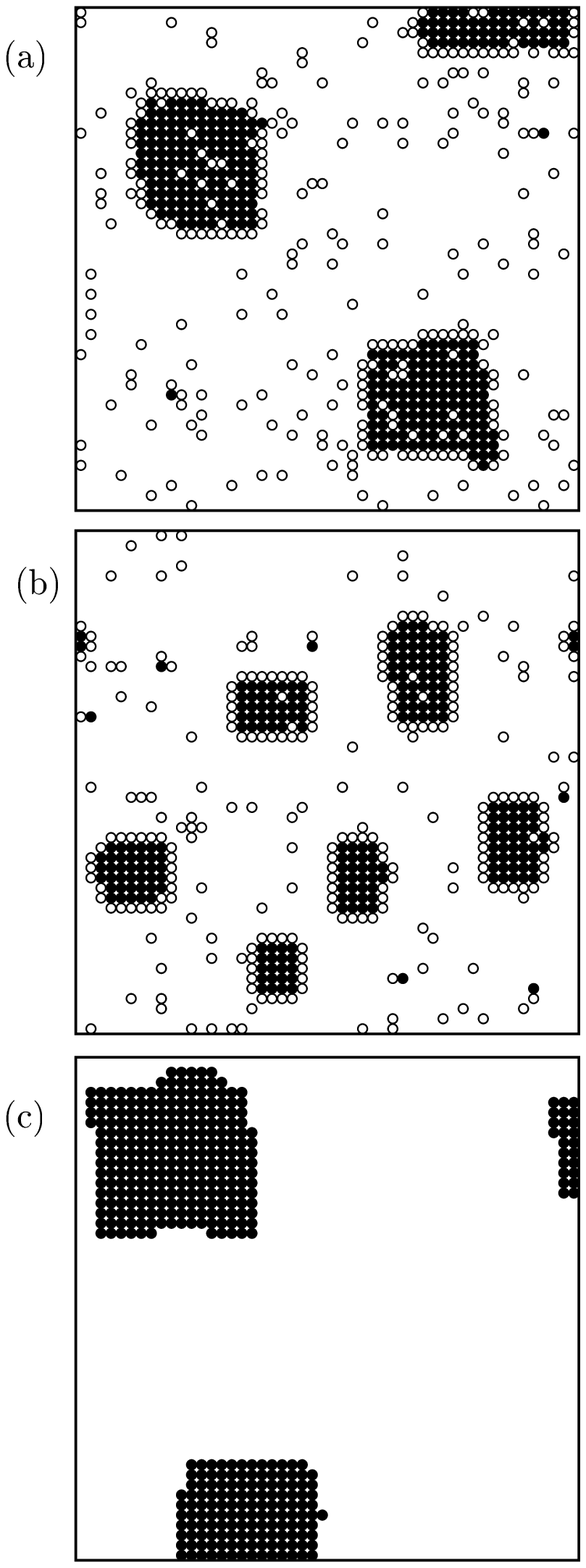}
\vspace*{70mm}
\caption{}
\label{fig:passivation}
\end{figure}

\newpage
\begin{figure}[hb]
\centering
\vspace*{140mm}
\includegraphics{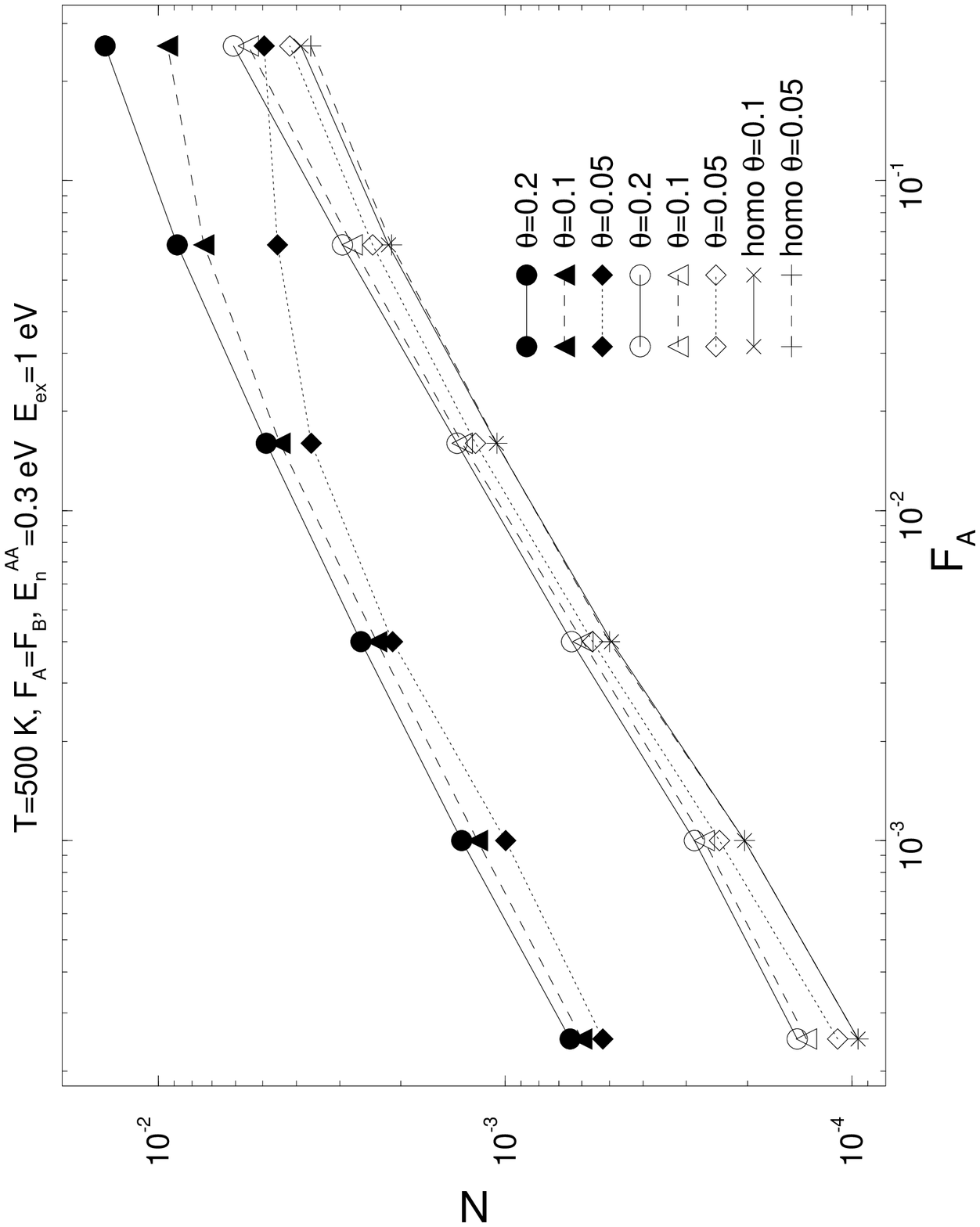}
\caption{}
\label{fig:var_par}
\end{figure}
\end{document}